\documentclass[11pt]{article}%
\usepackage{amsmath}
\usepackage{amsfonts}
\usepackage{amssymb}
\usepackage{graphicx}%
\providecommand{\U}[1]{\protect\rule{.1in}{.1in}}
\begin{document}

\title{Virial coefficients expressed by heat kernel coefficients}
\author{Xia-Qing Xu, Mi Xie\\{\footnotesize Department of Physics, Tianjin University, Tianjin 300072, P.
R. China}}
\date{}
\maketitle

\begin{abstract}
In this paper, we generally expressed the virial expansion of ideal quantum
gases by the heat kernel coefficients for the corresponding Laplace type
operator. As examples, we give the virial coefficients for quantum gases in
$d$-dimensional confined space and spheres, respectively. Our results show
that, the relative correction from the boundary to the second virial
coefficient is independent of the dimension and it always enhances the quantum
exchange interaction. In $d$-dimensional spheres, however, the influence of
the curvature enhances the quantum exchange interaction in two dimensions, but
weakens it in higher dimensions ($d>3$).

\end{abstract}

\section{Introduction}

Virial expansion is a powerful tool to study the effects of interaction on the
thermodynamic properties of many-particle systems
\cite{virial,virial2,virial3}. Virial expansion is first introduced to
describe weakly interacting classical gases. It expresses the equation of
state of the gas as a series of the number density, and the influence of
interaction is embodied in the higher-order terms. After the establishment of
quantum statistical mechanics, the virial expansion becomes even more
important. On the one hand, a quantum gas is usually treated in the grand
canonical ensemble in order to perform the sum over the states, but this
treatment introduces a non-independent parameter, the fugacity $z$, into the
equation of state. However, this parameter is eliminated in the virial
expansion, which leads to an equation of state without the extra parameter. On
the other hand, the virial expansion for a quantum gas directly reflects the
influence of quantum exchange interaction. Even for an ideal quantum gas in
which there is no classical interaction among the particles, the virial
expansion contains infinite number of terms, which reflect the influence of
the exchange symmetry of identical particles on the thermodynamic properties
of the gas. In other words, the virial expansion contains the contributions
not only from the classical interaction, but also from the quantum exchange interaction.

The heat kernel, as a kind of spectral functions, has close connections with
the spectrum asymptotics \cite{Kac} and other spectral functions, such as the
spectral zeta function \cite{zeta,zeta2} and the counting functions
\cite{DX2009} in mathematics. Since the first several heat kernel coefficients
can be analytically expressed in terms of geometric invariants of the manifold
\cite{Vassilevich,DeWitt,BG}, the asymptotic expansion of heat kernel, i.e.,
the heat kernel expansion, becomes a very important tool
\cite{Vassilevich,KirstenBK,Gilkey}. The heat kernel expansion has been
applied in many fields of physics, such as quantum field theory
\cite{QFT,QFT2,QFT3}, quantum gravity \cite{gravity,gravity2}, and string
theory \cite{AGM}. In quantum statistical mechanics, the sum over the spectrum
is a key component to calculate the grand partition function, so the heat
kernel approach also plays an important role \cite{KirstenBK,Avramidi,KT}.

In this paper, we will discuss the virial expansion of nonrelativistic ideal
quantum gases by using the approach of the heat kernel expansion. We will
first give the general expressions of the virial coefficients in terms of the
heat kernel coefficients. By these results, as long as the heat kernel
coefficients are known, the virial expansion of the equation of state of
quantum gases can be obtained straightforward. Then as examples, we will
calculate the virial expansions for quantum gases in $d$-dimensional confined
space and in $d$-dimensional sphere $S^{d}$, respectively. According to the
virial coefficients, we will analyze the influence of the boundary and the
curvature on the thermodynamic properties of quantum gases.

The paper is organized as follows. In section 2, we will give the general
results of the virial coefficients for ideal quantum gases expressed by the
global heat kernel and the heat kernel coefficients. In section 3, we will
take $d$-dimensional confined space and spheres as examples, and analyze the
influence of the boundary and the curvature on the virial expansion. The
conclusion is summarized in section 4. In Appendix A, the concrete results of
the virial coefficients for ideal Bose gases in $S^{d}$ are given.

\section{The relation between virial coefficients and heat kernel
coefficients}

In statistical mechanics, quantum gases are usually treated within the grand
canonical ensemble. For a nonrelativistic ideal Bose gas with a fixed particle
number, the equation of state can be expressed as%

\begin{align}
\frac{PV}{k_{B}T}  &  =\ln\Xi=-\sum_{n}\ln\left(  1-ze^{-\beta\varepsilon_{n}%
}\right)  ,\label{seq-1}\\
N  &  =\left(  z\frac{\partial}{\partial z}\ln\Xi\right)  _{V,T},
\label{seq-2}%
\end{align}
where $k_{B}$ is the Boltzmann constant, $\Xi$ is the grand partition
function, and $\varepsilon_{n}$ denotes the single-particle eigenenergies. For
clarity, in the following we will only consider Bose gases; at the end of this
paper, we will show how to obtain the Fermi results from the Bose ones. Since
the number of particles $N$ is a constant, the fugacity $z$ is not an
independent parameter and should be eliminated from the equation of state.
However, the explicit expression of the fugacity $z$ is hard to find out. Only
for some special cases, $z$ is solvable and the equation of state can be
expressed in an explicit form without $z$. Among them, the most important
approximate method is the virial expansion. When $z\ll1$, eqs. (\ref{seq-1})
and (\ref{seq-2}) can be expanded as serieses of $z$ so that a series solution
of $z$ can be obtained. Then the form of the equation of state becomes a
series of the number density of particles, which is the virial expansion.

According to eqs. (\ref{seq-1}) and (\ref{seq-2}), to obtain the equation of
state of a quantum gas, one needs to know the energy spectrum $\varepsilon
_{n}$ of the particles and to perform the sum over the spectrum in principle,
but this is usually very difficult. Fortunately, this difficulty can be
overcome by the help of the heat kernel approach. In the following we will
generally discuss how to achieve the virial expansion of the equation of state
in terms of the global heat kernel and the heat kernel expansion.

The global heat kernel of a Laplace type operator $D$ on a $d$-dimensional
manifold is defined as \cite{Vassilevich}%

\begin{equation}
K\left(  t\right)  =\sum_{n}e^{-\lambda_{n}t}, \label{Kt}%
\end{equation}
where $\left\{  \lambda_{n}\right\}  $ is the spectrum of the operator $D$.
When $t\ll1$, it can be expanded as a series, which is the heat kernel
expansion,
\begin{equation}
K\left(  t\right)  \approx\left(  4\pi t\right)  ^{-d/2}\sum_{k=0,\frac{1}%
{2},1,\cdots}^{\infty}B_{k}t^{k},\text{ \ \ \ }\left(  t\rightarrow0\right)
\label{HKexp}%
\end{equation}
where $B_{k}$ $\left(  k=0,1/2,1,\cdots\right)  $ are the heat kernel coefficients.

To deal with the sum in eq. (\ref{seq-1}), we first expand the logarithm term
to a series, then exchange the order of these two sums. Concretely, when
$z\ll1$, we have
\begin{equation}
\ln\Xi=\sum_{k=1}^{\infty}\frac{z^{k}}{k}\sum_{n}e^{-k\beta\varepsilon_{n}}.
\label{e1}%
\end{equation}
Since the eigenenergy $\varepsilon_{n}$ satisfies the equation%
\begin{equation}
\left[  -\nabla^{2}+\frac{2m}{\hbar^{2}}U\left(  \mathbf{x}\right)  \right]
\psi_{n}\left(  \mathbf{x}\right)  =\frac{2m}{\hbar^{2}}\varepsilon_{n}%
\psi_{n}\left(  \mathbf{x}\right)  , \label{Scheq}%
\end{equation}
the sum in eq. (\ref{e1}) can be represented by the global heat kernel for the
$D=-\nabla^{2}+\left(  2m/\hbar^{2}\right)  U\left(  \mathbf{x}\right)  $ as
\begin{equation}
\ln\Xi=\sum_{k=1}^{\infty}\frac{1}{k}K\left(  k\frac{\hbar^{2}\beta}%
{2m}\right)  z^{k}=\sum_{k=1}^{\infty}K\left(  k\frac{\lambda^{2}}{4\pi
}\right)  \frac{z^{k}}{k}, \label{e2}%
\end{equation}
where $\lambda=\hbar\sqrt{2\pi\beta/m}$ is the mean thermal wavelength.
Therefore, the grand potential can be expressed as a series of the global heat
kernels. Similarly, we can rewrite the particle number eq. (\ref{seq-2}) as%
\begin{equation}
N=\sum_{k=1}^{\infty}K\left(  k\frac{\lambda^{2}}{4\pi}\right)  z^{k},
\end{equation}
or%
\begin{equation}
n\lambda^{d}=\frac{\lambda^{d}}{V}\sum_{k=1}^{\infty}K\left(  k\frac
{\lambda^{2}}{4\pi}\right)  z^{k}, \label{nlambdad}%
\end{equation}
where $V$ is the volume of the system in $d$-dimensional space, and $n=N/V$ is
the number density of particles. To solve the asymptotic solution of the
fugacity $z$ from eq. (\ref{nlambdad}), assume that%
\begin{equation}
z=\sum_{k=1}^{\infty}c_{k}\left(  n\lambda^{d}\right)  ^{k}.\text{
\ \ \ }\left(  n\lambda^{d}\ll1\right)  \label{zexp}%
\end{equation}
Substituting it into eq. (\ref{nlambdad}) gives the coefficients $c_{k}$. Then
from eq. (\ref{e2}), we can obtain the virial expansion of the equation of
state as%
\begin{equation}
\frac{PV}{Nk_{B}T}=b_{1}+b_{2}n\lambda^{d}+b_{3}\left(  n\lambda^{d}\right)
^{2}+b_{4}\left(  n\lambda^{d}\right)  ^{3}+\cdots, \label{eq_state}%
\end{equation}
where the virial coefficients are
\begin{align}
b_{1}  &  =1,\nonumber\\
b_{2}  &  =-\frac{K\left(  2\frac{\lambda^{2}}{4\pi}\right)  }{2K^{2}\left(
\frac{\lambda^{2}}{4\pi}\right)  }\frac{V}{\lambda^{d}},\nonumber\\
b_{3}  &  =\left[  -\frac{2K\left(  3\frac{\lambda^{2}}{4\pi}\right)  }%
{3K^{3}\left(  \frac{\lambda^{2}}{4\pi}\right)  }+\frac{K^{2}\left(
2\frac{\lambda^{2}}{4\pi}\right)  }{K^{4}\left(  \frac{\lambda^{2}}{4\pi
}\right)  }\right]  \left(  \frac{V}{\lambda^{d}}\right)  ^{2},\nonumber\\
b_{4}  &  =\left[  -\frac{3K\left(  4\frac{\lambda^{2}}{4\pi}\right)  }%
{4K^{4}\left(  \frac{\lambda^{2}}{4\pi}\right)  }+\frac{3K\left(
2\frac{\lambda^{2}}{4\pi}\right)  K\left(  3\frac{\lambda^{2}}{4\pi}\right)
}{K^{5}\left(  \frac{\lambda^{2}}{4\pi}\right)  }-\frac{5K^{3}\left(
2\frac{\lambda^{2}}{4\pi}\right)  }{2K^{6}\left(  \frac{\lambda^{2}}{4\pi
}\right)  }\right]  \left(  \frac{V}{\lambda^{d}}\right)  ^{3},\nonumber\\
&  \vdots\label{bk_K}%
\end{align}
Thus we have expressed the virial coefficients for ideal Bose gases by the
global heat kernel for the corresponding operator.

On the other hand, the exact solutions of global heat kernels are not easy to
achieve, but the heat kernel expansion is easier to obtain. Expressing the
virial coefficients by heat kernel coefficients will provide a more powerful
way to perform the virial expansion. By use of the heat kernel expansion
(\ref{HKexp}), we can write%
\begin{equation}
K\left(  \frac{\lambda^{2}}{4\pi}\right)  \approx\sum_{k=0,\frac{1}%
{2},1,\cdots}^{\infty}\frac{B_{k}}{\left(  4\pi\right)  ^{k}}\lambda^{2k-d}.
\label{Kbeta}%
\end{equation}
Note that the heat kernel coefficient $B_{k}$ has a dimension $\left[
L\right]  ^{d-2k}$, so the sum in eq. (\ref{Kbeta}) can be regarded as a
series of $\lambda/\bar{L}$, where $\bar{L}$ represents the characteristic
length scale of the system. This result will be seen more clearly from the
specific examples in the next section.

Applying eq. (\ref{Kbeta}) to the equation of the grand potential eq.
(\ref{e2}), we arrive%

\begin{equation}
\ln\Xi\approx\sum_{k=0,\frac{1}{2},1,\cdots}^{\infty}\frac{B_{k}}{\left(
4\pi\right)  ^{k}}\lambda^{2k-d}g_{d/2+1-k}\left(  z\right)  ,\label{qexp}%
\end{equation}
where%
\begin{equation}
g_{\sigma}\left(  z\right)  =\frac{1}{\Gamma\left(  \sigma\right)  }\int
_{0}^{\infty}\frac{x^{\sigma-1}}{z^{-1}e^{x}-1}dx=\sum_{k=1}^{\infty}%
\frac{z^{k}}{k^{\sigma}}\label{BEexp}%
\end{equation}
is the Bose-Einstein integral. Similarly, the equation for particle number can
be rewritten as
\begin{equation}
n\lambda^{d}\approx\sum_{k=0,\frac{1}{2},1,\cdots}^{\infty}\frac{B_{k}%
}{\left(  4\pi\right)  ^{k}V}\lambda^{2k}g_{d/2-k}\left(  z\right)
.\label{nlambdadexp}%
\end{equation}
Eqs. (\ref{qexp}) and (\ref{nlambdadexp}) show that the asymptotic behavior of
an ideal Bose gas always takes the form of a series of the Bose-Einstein
integrals. This form of the equation of state has been used to discuss the
properties of quantum gases for decades \cite{KirstenBK,Avramidi,KT}. Since
the information of shape and topology of a system, as well as the external
potential, is embodied in the heat kernel coefficients, this approach is
widely applied to the quantum gases, especially the phenomenon of
Bose-Einstein condensation, in cavities \cite{KT,boundary,Xie}, in external
potentials \cite{Xie,KT2} and on different manifolds \cite{NT,FK}. For
electromagnetic fields, the heat kernel expansion is an effective tool to
study the Casimir effect for various geometries \cite{NLS,BKMM,Teo}. The
thermodynamical properties of relativistic quantum gases and the Bose-Einstein
condensation are also discussed in flat and curved space
\cite{Avramidi,AEGT,ADI}. However, there are few works to systematically
discuss the virial expansion in terms of the heat kernel expansion.

Taking advantaging of the expansion of the Bose-Einstein integral eq.
(\ref{BEexp}) and following the same procedure as above, we can obtain the
final result of the virial coefficients expressed by the heat kernel
coefficients. However, the complete expressions are too complicated to present
here, so we will divide the problem into two parts according to whether the
heat kernel expansion contains half-integer power terms or not.

When the heat kernel expansion eq. (\ref{HKexp}) contains half-integer power
terms, the first several virial coefficients take the forms as
\begin{align}
b_{1}  &  =1,\nonumber\\
b_{2}  &  \approx-\frac{1}{2^{d/2+1}}+\frac{\sqrt{2}-1}{2^{\left(  d+3\right)
/2}}\frac{\lambda B_{1/2}}{\sqrt{\pi}V}-\frac{3-2\sqrt{2}}{2^{\left(
d+6\right)  /2}}\left(  \frac{\lambda B_{1/2}}{\sqrt{\pi}V}\right)
^{2},\nonumber\\
b_{3}  &  \approx-\left(  \frac{2}{3^{d/2+1}}-\frac{1}{2^{d}}\right)  +\left(
-\frac{\sqrt{2}-1}{2^{d-1/2}}+\frac{\sqrt{3}-1}{3^{\left(  d+1\right)  /2}%
}\right)  \frac{\lambda B_{1/2}}{\sqrt{\pi}V}+\left(  \frac{3-2\sqrt{2}}%
{2^{d}}-\frac{2-\sqrt{3}}{2\cdot3^{d/2}}\right)  \left(  \frac{\lambda
B_{1/2}}{\sqrt{\pi}V}\right)  ^{2},\nonumber\\
b_{4}  &  \approx-\left(  \frac{3}{4^{d/2+1}}-\frac{1}{2^{d/2}\cdot3^{d/2-1}%
}+\frac{5}{2^{3d/2+1}}\right)  +\left(  \frac{3}{2^{d+2}}+\frac{15\left(
\sqrt{2}-1\right)  }{2^{\left(  3d+3\right)  /2}}-\frac{5\sqrt{2}-\sqrt{6}%
-2}{2^{\left(  d+3\right)  /2}3^{d/2-1}}\right)  \frac{\lambda B_{1/2}}%
{\sqrt{\pi}V}\nonumber\\
&  +\left[  -\frac{3}{2^{d+3}}-\frac{45\left(  3-2\sqrt{2}\right)
}{2^{3d/2+3}}+\frac{15\sqrt{2}-10+2\sqrt{3}-5\sqrt{6}}{2^{\left(  d+5\right)
/2}3^{d/2-1}}\right]  \left(  \frac{\lambda B_{1/2}}{\sqrt{\pi}V}\right)
^{2}, \label{virial_1/2}%
\end{align}
where we have used the relation $B_{0}=V$. These expressions are accurate to
the second order in $\lambda$. From eq. (\ref{virial_1/2}), we can find that
both the terms proportional to $\lambda^{1}$ and $\lambda^{2}$ are only
related to the coefficient $B_{1/2}$. Other higher-order heat kernel
coefficients only contribute higher-power terms of $\lambda$.

When the heat kernel expansion only contains integer power terms, the virial
coefficients are%
\begin{align}
b_{1}  &  =1,\nonumber\\
b_{2}  &  \approx-\frac{1}{2^{d/2+1}}+\frac{1}{2^{d/2+5}}\frac{\lambda
^{4}\left(  B_{1}^{2}-2VB_{2}\right)  }{\pi^{2}V^{2}},\nonumber\\
b_{3}  &  \approx-\left(  \frac{2}{3^{d/2+1}}-\frac{1}{4^{d/2}}\right)
+\left(  \frac{1}{2^{3}\cdot3^{d/2}}-\frac{1}{2^{d+3}}\right)  \frac
{\lambda^{4}\left(  B_{1}^{2}-2VB_{2}\right)  }{\pi^{2}V^{2}},\nonumber\\
b_{4}  &  \approx-\left(  \frac{3}{4^{d/2+1}}-\frac{1}{2^{d/2}3^{d/2-1}}%
+\frac{5}{2^{3d/2+1}}\right)  +\left(  \frac{9}{2^{d+5}}-\frac{1}%
{2^{d/2+2}3^{d/2-1}}+\frac{15}{2^{3d/2+5}}\right)  \frac{\lambda^{4}\left(
B_{1}^{2}-2VB_{2}\right)  }{\pi^{2}V^{2}}, \label{virail_1}%
\end{align}
where we have also taken $B_{0}=V$. In these expressions we only include the
leading correction terms which are proportional to $\lambda^{4}$. Clearly, the
leading correction to the virial coefficients is related to both the heat
kernel coefficients $B_{1}$ and $B_{2}$. Furthermore, the leading correction
to each virial coefficient (except $b_{1}$) has a similar form. They are all
proportional to the same factor $\lambda^{4}\left(  B_{1}^{2}-2VB_{2}\right)
/V^{2}$, only with different numeric constants.

In the following, we will consider two specific cases as the examples of the
above results.

\section{Examples}

\subsection{$d$-dimensional confined space}

A thermodynamic system is usually assumed to be extensive, which implies that
the effect of boundary is neglected. For a macro system, this is a good
approximation, but for a small-scale system, the influence of boundary becomes
large and may not be ignored. For example, when considering quantum dots or
quantum wires with dimensions in the nanometer range, the effects of boundary
must be included. In recent years, much research on the influence of boundary
on the thermodynamic properties has been carried out
\cite{boundary,boundary2,boundary3,boundary4,boundary5}.

Now we will consider the influence of boundary on the virial expansion of an
ideal Bose gas by use of the approach given in the above section. On a
manifold with boundary, the heat kernel expansion of the Laplace operator has
half-integer power terms. For the Dirichlet boundary conditions, only taking
the first two terms, the heat kernel expansion has the general form
\cite{Vassilevich}%
\begin{equation}
K\left(  t\right)  \approx\frac{1}{\left(  4\pi t\right)  ^{d/2}}\left(
V-\frac{\sqrt{\pi}}{2}St^{1/2}\right)  ,
\end{equation}
where $V$ and $S$ represent the volume and surface area of the $d$-dimensional
manifold, respectively. In other words, the second heat kernel coefficient has
the general form $B_{1/2}=-\left(  \sqrt{\pi}/2\right)  S$ in arbitrary
dimensions. Substituting it into eq. (\ref{virial_1/2}) immediately gives the
virial coefficients of an ideal Bose gas in $d$-dimensional confined space.

Clearly, the most important correction to the thermodynamic properties of the
quantum gas is described by the second virial coefficient $b_{2}$. In confined
space, the main contribution of the boundary is embodied in the correction to
$b_{2}$. To describe the correction, we introduce the relative correction
parameter $\eta$ as the ratio of the correction to $b_{2}$ and its zero-order
term. From eq. (\ref{virial_1/2}), we find
\begin{equation}
\eta=\frac{2-\sqrt{2}}{4}\frac{S\lambda}{V}+\frac{3-2\sqrt{2}}{16}\left(
\frac{S\lambda}{V}\right)  ^{2}. \label{eta_B}%
\end{equation}
Therefore, the relative correction is only related to the factor $S\lambda/V$
but is independent of the dimension. It indicates that regardless of the
dimension of the system, the influence of the boundary on the thermodynamic
properties of quantum gases is almost the same.

In a roughly isotropic system, since $V/S$ represents the length scale of the
system $\bar{L}$, the relative correction $\eta$ given in eq. (\ref{eta_B})
only related to the ratio of the mean wavelength and the scale of the system
$\lambda/\bar{L}$. This is consistent with the general analysis about the
expansion parameter of the heat kernel expansion in last section. It indicates
that eq. (\ref{eta_B}) holds for small values of $\lambda/\bar{L}$. As a
result, the influence of the boundary will become manifest for a small-scale
system under a relatively low temperature, in which the mean thermal
wavelength of particles is not much smaller than the scale of the system and
$\lambda/\bar{L}$ is not negligible. For example, in a cube with side length
$L$, when the temperature satisfies $\lambda/L=0.1$, the influence of boundary
to the second virial coefficient $b_{2}$ is $\eta=9.2\%$ according to eq.
(\ref{eta_B}). On the other hand, in an highly anisotropic system, e.g., a
long and narrow one, the ratio $S/V$ is larger, so the influence of boundary
is more important in such cases.

\subsection{$d$-dimensional sphere}

In this section, we will consider the virial expansion of Bose gases when the
corresponding heat kernel expansion only contains integer terms. A typical
example is the heat kernel of the Laplace operator on a smooth curved
manifold. For simplification, in the following we will consider the case of
$d$-dimensional sphere $S^{d}$, which is a typical representative of constant
curvature spaces.

We denote the sectional curvature of the space as $\kappa$. The heat kernel in
$S^{d}$ can be generally expressed as \cite{Nagase}%
\begin{align}
K_{2k+1}\left(  \rho,t\right)   &  =\frac{\left(  -1\right)  ^{k}}{2^{k}%
\pi^{k}}\frac{1}{\sqrt{4\pi t}}\left(  \frac{\sqrt{\kappa}}{\sin\left(
\sqrt{\kappa}\rho\right)  }\frac{\partial}{\partial\rho}\right)  ^{k}%
e^{k^{2}\kappa t-\frac{\rho^{2}}{4t}},\nonumber\\
K_{2k+2}\left(  \rho,t\right)   &  =\frac{\left(  -1\right)  ^{k}}%
{2^{k+5/2}\pi^{k+3/2}t^{3/2}}e^{\frac{\left(  2k+1\right)  ^{2}}{4}\kappa
t}\left(  \frac{\sqrt{\kappa}}{\sin\left(  \sqrt{\kappa}\rho\right)  }%
\frac{\partial}{\partial\rho}\right)  ^{k}\int_{\rho}^{\delta/\sqrt{\left\vert
\kappa\right\vert }}\frac{\sqrt{\kappa}se^{-\frac{s^{2}}{4t}}}{\sqrt
{\cos\left(  \sqrt{\kappa}\rho\right)  -\cos\left(  \sqrt{\kappa}s\right)  }%
}ds,\nonumber\\
&  \left(  k=0,1,2,\cdots\right)  \label{Kt_local}%
\end{align}
where $\rho$ is the geodesic distance and $0<\delta\,<\pi/2$. Taking trace of
the heat kernel and after some calculations, we can obtain the following
expressions for the global heat kernel for different dimensions.%
\begin{align}
K_{2}\left(  t\right)   &  =V\frac{e^{\frac{1}{4}\kappa t}}{4\pi t}\left[
1+\sum_{l=1}^{\infty}\left(  2^{1-2l}-1\right)  B_{2l}\frac{\left(  -\kappa
t\right)  ^{l}}{l!}\right]  ,\nonumber\\
K_{3}\left(  t\right)   &  =V\frac{e^{\kappa t}}{\left(  4\pi t\right)
^{3/2}},\nonumber\\
K_{4}\left(  t\right)   &  =V\frac{e^{\frac{9}{4}\kappa t}}{\left(  4\pi
t\right)  ^{2}}\left\{  1-\frac{\kappa t}{4}+\sum_{l=2}^{\infty}\frac{\left(
-\kappa t\right)  ^{l}}{l!}\left[  -\left(  l-1\right)  \left(  2^{1-2l}%
-1\right)  B_{2l}+\frac{l}{4}\left(  2^{3-2l}-1\right)  B_{2\left(
l-1\right)  }\right]  \right\}  ,\nonumber\\
K_{5}\left(  t\right)   &  =V\frac{e^{4\kappa t}}{\left(  4\pi t\right)
^{5/2}}\left(  1-\frac{2}{3}\kappa t\right)  . \label{Kt_d}%
\end{align}
These results show a difference between odd and even dimensions. The
expression for the odd dimension only contains finite number of terms but that
for the even dimension contains infinite terms, which is a natural result led
by the heat kernel for different dimensions, eq. (\ref{Kt_local}).

In the expressions (\ref{Kt_d}), the $3$-d case is special since the global
heat kernel $K_{3}\left(  t\right)  $ differs to that in flat $3$-d space only
by an exponential factor $e^{\kappa t}$. One can generally prove that when the
only difference between two heat kernels is an exponential factor, i.e.,
$K^{\left(  2\right)  }\left(  t\right)  =K^{\left(  1\right)  }\left(
t\right)  e^{\xi t}$, the virial coefficients related to them are the same.
This result can be directly verified from eq. (\ref{bk_K}). Therefore, the
virial expansion of ideal quantum gases in a $3$-d sphere $S^{3}$ is exactly
the same as that in $3$-d flat space. Notice that a trivial case is that the
spectra of two systems only differ by a translation, which corresponds a
different choice of the zero-point energy, and this certainly does not affect
the statistical or thermodynamic properties of the system. Nevertheless, the
above example is different. The spectrum of the Laplace operator in a $3$-d
sphere has a different structure from that in $3$-d flat space. The
equivalence of these systems is not trivial.

Further expanding the exponential factor in eq. (\ref{Kt_d}) will give the
heat kernel expansion. Then the virial coefficients in the corresponding space
can be obtained directly. We list the virial coefficients in different
dimensions in Appendix A.

As discusses before, the influence of the curvature on the quantum gas is
mainly reflected by the relative correction $\eta$, which is the proportion of
the correction to $b_{2}$. From the results given in Appendix A, we can find
that, different from the case of confined space, the relative correction in
$S^{d}$ is tightly related the dimension. The value of $\eta$ in different
dimensions are listed in Table 1, in which we only consider the leading contribution.

\begin{table}[ptb]
\begin{center}%
\begin{tabular}
[c]{cccccccccc}\hline\hline
dimension & $2$ & $3$ & $4$ & $5$ & $6$ & $7$ & $8$ & $9$ & $10$\\\hline\hline
$\eta\left(  \frac{\kappa^{2}\lambda^{4}}{\pi^{2}}\right)  $ & $\frac{1}{720}$
& $0$ & $-\frac{1}{120}$ & $-\frac{1}{36}$ & $-\frac{1}{16}$ & $-\frac{7}{60}
$ & $-\frac{7}{36}$ & $-\frac{3}{10}$ & $-\frac{7}{16}$\\\hline
\end{tabular}
\end{center}
\caption{The leading relative correction $\eta$ in $d$-dimensional sphere.}%
\end{table}

As shown in Table 1, the leading correction to the virial coefficient
$\eta\sim\left(  \lambda/R\right)  ^{4}$, where $R$ is the radius of the
$d$-sphere and it also represents the length scale of the system, which is
consistent with the analysis of the heat kernel expansion in last section.
From Table 1, we can find the two-dimensional case is different from others.
The relative correction $\eta$ in two dimensions is positive, which means that
the influence of the curved space has the same sign as the quantum exchange
interaction, or, the curved space enhances the quantum exchange interaction.
However, in the space with $d>3$, the sign of $\eta$ becomes negative, which
means that the influence of the curvature will weaken the quantum exchange
interaction, and this influence becomes larger with the increase of the
dimension. The three-dimensional space is the critical case, in which the
correction vanishes, just as the above analysis.

\section{Discussion and Conclusion}

In the above sections we only consider the Bose gases for clarity. As for
Fermi gases, the treatment is exactly the same, and the results only have some
minor differences: Besides the grand potential (\ref{seq-1}) changes to
$\ln\Xi=\sum_{n}\ln\left(  1+ze^{-\beta\varepsilon_{n}}\right)  $ and the
Bose-Einstein integrals in eqs. (\ref{qexp}) and (\ref{nlambdadexp}) change to
the Fermi-Dirac integrals, the main difference is that all of the even-order
virial coefficients $b_{2},b_{4},\cdots$ will have an extra negative sign
while the odd-order ones $b_{1},b_{3},\cdots$ are the same as the Bose case.
Therefore, the virial expansion for Fermi gases can be obtained from the above
Bose results directly.

In this paper we focus on the virial expansion, which is in principle
applicable to the high temperature and low density region. The applicability
range is often represented as $n\lambda^{3}\ll1$. However, we know that the
virial expansion usually can be applied to a wider range of the parameter.
Taking the $3$-d flat space case as an example, in which the virial expansion
for an ideal Bose gas is%
\begin{equation}
\frac{PV}{Nk_{B}T}\approx1-0.1768n\lambda^{3}-0.0033\left(  n\lambda
^{3}\right)  ^{2}-0.00011\left(  n\lambda^{3}\right)  ^{3}+\cdots.
\end{equation}
Due to the rapid decrease of the virial coefficients, the approximation is
usually quite good. For example, even for $n\lambda^{3}\sim1$, only taking
into account the first two terms gives an error about $0.3\%$; if taking first
three terms, the error is about $0.01\%$. As a comparison, the Bose-Einstein
condensation for a free ideal gas occurs at $n\lambda^{3}\approx2.612$, which
implies that the virial expansion is a good approximation as long as the
temperature is not very closed to the critical temperature. In Ref. \cite{DP},
the authors detailed analyze the applicability range of the virial expansion
for relativistic pion gases and reach a similar conclusion.

Our results can be verified by experiments in small-scale systems. Since a
real system exists in no more than three dimensions, in our results, those can
be directly compared with experiments mainly includes two- or
three-dimensional systems with boundaries and two-dimensional spheres with the
length scales in the nanometer range. In experiments, the effect of boundary
can be found for non-interacting real particles or quasiparticles, e.g.,
electrons or excitons, in some nanoparticles and nanofilms at an appropriate
temperature. Our results show that the influence of boundary becomes stronger
in highly anisotropic systems. The influence of curvature is expected to be
observed in some systems with cage shapes, e.g., electrons in a fullerene, or
quasiparticles on the surface of a spherical system, e.g., surface excitons on
the surface of xenon clusters \cite{DDK}.

In summary, in this paper, we discuss the relation between the virial
coefficients of an ideal quantum gas and the heat kernel coefficients for the
corresponding Laplace type operator, and give the expressions of the virial
coefficients expressed by the heat kernel coefficients. As examples, we
discuss the quantum gases in confined space and in spheres in any dimensions,
and give the explicit results for the virial expansions. Our results show
that, in confined space, the influence of boundary is almost independent of
the dimension of space, and its effect enhances the quantum exchange
interaction of the gas. On the other hand, the influence of curvature on
quantum gases is tightly related to the dimension of space. It can enhance
($d=2$) or weaken ($d>3$) the effect of the quantum exchange interaction.

\section*{Acknowledgement}

This work is supported in part by NSF of China under Project No. 11575125.

\section*{Appendix A}

The first four virial coefficients for ideal Bose gases in $d$-dimensional
spheres $S^{d}$ are

(1) $2$-d:%
\begin{align*}
b_{1}  &  =1,\\
b_{2}  &  \approx-\frac{1}{4}-\frac{\kappa^{2}\lambda^{4}}{2880\pi^{2}}%
-\frac{\kappa^{3}\lambda^{6}}{15120\pi^{3}}-\frac{\kappa^{4}\lambda^{8}%
}{69120\pi^{4}},\\
b_{3}  &  \approx\frac{1}{36}-\frac{\kappa^{2}\lambda^{4}}{4320\pi^{2}}%
-\frac{\kappa^{3}\lambda^{6}}{9720\pi^{3}}-\frac{1871\kappa^{4}\lambda^{8}%
}{43545600\pi^{4}},\\
b_{4}  &  \approx-\frac{\kappa^{2}\lambda^{4}}{11520\pi^{2}}-\frac{\kappa
^{3}\lambda^{6}}{12096\pi^{3}}-\frac{29\kappa^{4}\lambda^{8}}{483840\pi^{4}}.
\end{align*}

(2) $3$-d:%
\begin{align*}
b_{1}  &  =1,\\
b_{2}  &  =-\frac{1}{4\sqrt{2}},\\
b_{3}  &  =-\left(  \frac{2}{9\sqrt{3}}-\frac{1}{8}\right)  ,\\
b_{4}  &  =-\left(  \frac{5}{32\sqrt{2}}-\frac{1}{2\sqrt{6}}+\frac{3}%
{32}\right)  ,
\end{align*}
which are the same as those in flat space.

(3) $4$-d:%
\begin{align*}
b_{1}  &  =1,\\
b_{2}  &  \approx-\frac{1}{8}+\frac{\kappa^{2}\lambda^{4}}{960\pi^{2}}%
+\frac{\kappa^{3}\lambda^{6}}{3360\pi^{3}}+\frac{7\kappa^{4}\lambda^{8}%
}{92160\pi^{4}}+\frac{61\kappa^{5}\lambda^{10}}{2956800\pi^{5}},\\
b_{3}  &  \approx-\frac{5}{432}+\frac{7\kappa^{2}\lambda^{4}}{8640\pi^{2}%
}+\frac{37\kappa^{3}\lambda^{6}}{90720\pi^{3}}+\frac{989\kappa^{4}\lambda^{8}%
}{5806080\pi^{4}},\\
b_{4}  &  \approx-\frac{1}{384}+\frac{5\kappa^{2}\lambda^{4}}{9216\pi^{2}%
}+\frac{13\kappa^{3}\lambda^{6}}{32256\pi^{3}}.
\end{align*}

(4) $5$-d:%
\begin{align*}
b_{1}  &  =1,\\
b_{2}  &  \approx-\frac{1}{8\sqrt{2}}+\frac{\kappa^{2}\lambda^{4}}{288\sqrt
{2}\pi^{2}}+\frac{\kappa^{3}\lambda^{6}}{864\sqrt{2}\pi^{3}}+\frac{\kappa
^{4}\lambda^{8}}{3456\sqrt{2}\pi^{4}}+\frac{\kappa^{5}\lambda^{10}}%
{15552\sqrt{2}\pi^{5}},\\
b_{3}  &  \approx\frac{1}{32}-\frac{2}{27\sqrt{3}}+\left(  \frac{1}%
{162\sqrt{3}}-\frac{1}{576}\right)  \frac{\kappa^{2}\lambda^{4}}{\pi^{2}%
}-\left(  \frac{1}{1728}-\frac{2}{729\sqrt{3}}\right)  \frac{\kappa^{3}%
\lambda^{6}}{\pi^{3}}+\left(  \frac{5}{5832\sqrt{3}}-\frac{5}{41472}\right)
\frac{\kappa^{4}\lambda^{8}}{\pi^{4}},\\
b_{4}  &  \approx-\left(  \frac{3}{128}+\frac{5}{256\sqrt{2}}-\frac{1}%
{12\sqrt{6}}\right)  +\left(  -\frac{1}{108\sqrt{6}}+\frac{5}{3072\sqrt{2}%
}+\frac{1}{256}\right)  \frac{\kappa^{2}\lambda^{4}}{\pi^{2}}.
\end{align*}

\end{document}